\documentclass[amsmath,amssymb,aps,prc,preprint,superscriptaddress]{revtex4-1}
\usepackage{graphicx}
\usepackage{longtable}
\usepackage{tabularx}
\usepackage{dcolumn}
\usepackage{xcolor}
\usepackage{bm}
\usepackage{booktabs}

\usepackage{CJK}

\begin{document}
\begin{CJK*}{GB}{} 

\title{Study of the $^{34}$Ar$(\alpha,p)^{37}$K reaction rate via proton scattering on $^{37}$K, and its impact on properties of modeled X-Ray bursts}




\author{A. Lauer-Coles}

\altaffiliation[Present address: ]{Savannah River National Laboratory, Savannah River Site, SC 29803, USA}
\homepage[]{www.amberlauercoles.com}
\email[Contact author:]{ amber.coles@srnl.doe.gov}

\author{C. M. Deibel}
\email[Contact author: ]{deibel@lsu.edu}
\affiliation{Louisiana State University Dept. of Physics 
\& Astronomy, Baton Rouge, LA 70803, USA}

\author{J. C. Blackmon}
\affiliation{Louisiana State University Dept. of Physics 
\& Astronomy, Baton Rouge, LA 70803, USA}

\author{S. Ahn}
\altaffiliation[Present address: ]{Center for Exotic Nuclear Studies, Institute for Basic Science, Daejeon, South Korea 34126}
\affiliation{Department of Physics and Astronomy and National Superconducting Cyclotron Laboratory, Michigan State University, East Lansing, MI 48824, USA}

\author{M. Anastasiou}
\altaffiliation[Present address: ]{Nuclear and Chemical Sciences Division, Lawrence Livermore National Laboratory, Livermore, CA 94550, USA}
\affiliation{Physics Department, Florida State University, Tallahassee, FL 32306, USA}

\author{L. T. Baby}
\affiliation{Physics Department, Florida State University, Tallahassee, FL 32306, USA}

\author{S. Bedoor}

\affiliation{Cyclotron Institute, Texas A\&M University, College Station, TX 77483, USA}

\author{J. Browne}
\affiliation{Department of Physics and Astronomy and National Superconducting Cyclotron Laboratory, Michigan State University, East Lansing, MI 48824, USA}

\author{K. A. Chipps}
\affiliation{Physics Division, Oak Ridge National Laboratory, Oak Ridge, TN 37831, USA}
\affiliation{Department of Physics and Astronomy, University of Tennessee, Knoxville, TN 37996, USA}

\author{E. C. Good}
\altaffiliation[Present address: ]{Pacific Northwest National Laboratory, 902 Battelle Blvd, Richland, WA 99354, USA}
\affiliation{Louisiana State University Dept. of Physics 
\& Astronomy, Baton Rouge, LA 70803, USA}

\author{A. Hood}
\affiliation{Louisiana State University Dept. of Physics 
\& Astronomy, Baton Rouge, LA 70803, USA}

\author{J. Hooker}
\altaffiliation[Present address: ]{Los Alamos National Laboratory, Los Alamos, NM 87545, USA}
\affiliation{Department of Physics and Astronomy, Texas A\&M University, College Station, TX 77843, USA}
\affiliation{Cyclotron Institute, Texas A\&M University, College Station, TX 77483, USA}

\author{H. Jayatissa}
\altaffiliation[Present address: ]{Los Alamos National Laboratory, Los Alamos, NM 87545, USA}
\affiliation{Department of Physics and Astronomy, Texas A\&M University, College Station, TX 77843, USA}
\affiliation{Cyclotron Institute, Texas A\&M University, College Station, TX 77483, USA}

\author{E. Koshchiy}
\altaffiliation[]{Deceased.}
\affiliation{Cyclotron Institute, Texas A\&M University, College Station, TX 77483, USA}

\author{K. T. Macon}
\affiliation{Louisiana State University Dept. of Physics 
\& Astronomy, Baton Rouge, LA 70803, USA}

\author{F. Montes}
\affiliation{Department of Physics and Astronomy and National Superconducting Cyclotron Laboratory, Michigan State University, East Lansing, MI 48824, USA}

\author{W. J. Ong}
\altaffiliation[Present address: ]{Nuclear and Chemical Sciences Division, Lawrence Livermore National Laboratory, Livermore, CA 94550, USA}
\affiliation{Department of Physics and Astronomy and National Superconducting Cyclotron Laboratory, Michigan State University, East Lansing, MI 48824, USA}

\author{S. D. Pain}
\affiliation{Physics Division, Oak Ridge National Laboratory, Oak Ridge, TN 37831, USA}
\affiliation{Department of Physics and Astronomy, University of Tennessee, Knoxville, TN 37996, USA}

\author{N. Rijal}
\affiliation{Physics Department, Florida State University, Tallahassee, FL 32306, USA}

\author{G. V. Rogachev}
\affiliation{Department of Physics and Astronomy, Texas A\&M University, College Station, TX 77843, USA}
\affiliation{Cyclotron Institute, Texas A\&M University, College Station, TX 77483, USA}

\author{D. Santiago-Gonzalez}
\affiliation{Argonne National Laboratory, Lemont, IL 60439, USA}

\author{H. Schatz}
\affiliation{Department of Physics and Astronomy and National Superconducting Cyclotron Laboratory, Michigan State University, East Lansing, MI 48824, USA}

\author{K. Schmidt}
\affiliation{Department of Physics and Astronomy and National Superconducting Cyclotron Laboratory, Michigan State University, East Lansing, MI 48824, USA}
\affiliation{Institute of Radiation Physics, Helmholtz-Zentrum Dresden-Rossendorf, Bautzner Landstr. 400, 01328 Dresden, Germany}

\author{S. Upadhyayula}
\altaffiliation[Present address: ]{Lawrence Livermore National Laboratory, Livermore, CA 94550, USA}
\affiliation{Department of Physics and Astronomy, Texas A\&M University, College Station, TX 77843, USA}
\affiliation{Cyclotron Institute, Texas A\&M University, College Station, TX 77483, USA}

\author{I. Wiedenh\"{o}ver}
\affiliation{Physics Department, Florida State University, Tallahassee, FL 32306, USA}



\pacs{}



\date{\today}
\begin{abstract}
\begin{description}
\item[Background] Type I X-Ray bursts (XRBs) are energetic stellar explosions that occur on the surface of a neutron star in an accreting binary system with a low-mass H/He-rich companion.  The rate of the $^{34}$Ar$(\alpha, p)^{37}$K reaction may influence features of the light curve that results from the underlying thermonuclear runaway, as shown in recent XRB stellar modelling studies. 

\item[Purpose] In order to reduce the uncertainty of the rate of this reaction, properties of resonances in the compound nucleus $^{38}$Ca, such as resonance energies, spins, and particle widths, must be well constrained.
\item[Method] This work discusses a study of resonances in the $^{38}$Ca compound nucleus produced in the $^{34}$Ar$(\alpha, p)$ reaction. The experiment was performed at the National Superconducting Cyclotron Laboratory, with the ReA3 facility by measuring proton scattering using an unstable $^{37}$K beam. The kinematics were designed specifically to identify and characterize resonances in the Gamow energy window for the temperature regime relevant to XRBs.

\item[Results] The spins and proton widths of newly identified and previously known states in $^{38}$Ca in the energy region of interest for the $^{34}$Ar$(\alpha, p)^{37}$K reaction have been constrained through an R-Matrix analysis of the scattering data.
\item[Conclusions] Using these constraints, a newly estimated rate is applied to an XRB model built using Modules for Experiments in Stellar Astrophysics (MESA), to examine its impact on observables, including the light curve.  It is found that the newly determined reaction rate does not substantially affect the features of the light curve. 
\end{description}
\end{abstract}

\maketitle
\end{CJK*}

\section{Introduction}\label{sec:intro}

Type I X-Ray Burst (XRB) nucleosynthesis occurs when H- and He-rich material is accreted onto the surface of a neutron star. This happens when the neutron star is in a low-mass X-ray binary system with a hydrogen- and/or helium-rich donor, such as a main sequence star. The surface material then seeds the resulting nuclear reactions, which first consist of nuclear burning that occurs on the surface during accretion through the Hot Carbon-Nitrogen-Oxygen (HCNO) cycles, and then the thermonuclear runaway during the XRB. This runaway occurs when the temperature-dependent triple-$\alpha$ reaction rate increases, significantly contributing energy to the system and leading to a thin-shell instability \cite{hansen_steady-state_1975}. In this mildly degenerate shell,  thermal energy from nucleosynthesis is not dissipated, and instead feeds back into the rising temperature, resulting in a shell-flash. The rising temperature allows additional nuclear reactions that lead to breakout of the hCNO cycles and subsequent nucleosynthesis via the $\alpha,p$ and $rp$ processes giving rise to the X-ray burst.  The burst is identified by the primary observable, a release of photons in the X-ray spectrum, known as the ``light-curve", characterized by a rapid increase in flux, followed by an exponential decay.  Type I XRBs can last $10-100$ s, reach peak temperatures of $1-2$ GK in the nuclear burning layer,  recur over periods of hours to days, and release approximately $10^{39}-10^{40}$ ergs of energy. 

The XRB is fueled during the initial rise time by the $(\alpha,p)$ process, a series of $(\alpha,p)$ and $(p,\gamma)$ reactions, followed by the exponential decay of the burst driven by the $rp$ (rapid proton-capture) process, a series of $(p,\gamma)$ reactions and $\beta$ decays ending around A $=100$ \cite{schatz_x-ray_2006-1}.  
 Several of the $(\alpha,p$) reactions in the former process occur on so-called waiting-point nuclei. In these even-even, $(Z-N)/2=1$ nuclei the radiative proton captures and the corresponding photo-disintegration are in a $(p,\gamma)-(\gamma,p)$ equilibrium, due to their low $Q_{p,\gamma}$-values. Without an alternate pathway, nucleosynthesis may stall due to the relatively long $\beta^+$-decay half-life ($\sim$1 s), which is comparable to the timescale of the burst. An $(\alpha,p)$ reaction, once the temperature is sufficiently high, offers an alternative route for the nucleosynthesis, thus restarting the burst. Such a stalling and restarting of the nucleosynthesis may be responsible for observed double-peaked XRB light curves (e.g. \cite{Pennix1989}).  There are four common candidates for the dominant waiting point nuclei in the $(\alpha,p)$ process, $^{22}$Mg, $^{26}$Si, $^{30}$S, and $^ {34}$Ar \cite{fisker_nuclear_2004}, and an additional possible candidate, $^{38}$Ca, identified in Ref. \cite{obrien_exploring_2009}. In standard, single-peaked bursts, the waiting point nuclei have also been shown to impact burst features. 

Several published stellar models
indeed indicate that the $^{34}$Ar$(\alpha,p)^{37}$K rate influences observables and other relevant quantities \cite{cyburt_dependence_2016,fisker_nuclear_2004, parikh_effects_2008}. Specifically, Fisker \textit{et al.} \cite{fisker_nuclear_2004} showed that $^{34}$Ar($\alpha,p$)$^{37}$K may affect the magnitude of the dip in a double-peaked burst. Cyburt \textit{et al.} \cite{cyburt_dependence_2016} in a 2016 XRB sensitivity study showed that this reaction may also influence the burst luminosity. Finally, in Parikh \textit{et al.} \cite{parikh_effects_2008}, variation in the rate was found to affect the $^{34}$S abundance. 

Due to low cross sections of the reaction and the low intensities of available $^{34}$Ar beams, the $^{34}$Ar($\alpha,p$)$^{37}$K reaction has only been studied directly once by Browne \textit{et al.} \cite{Browne_2023}, though at energies well above the astrophysically relevant energy regime. It has yet to be incorporated into rate formulations in the standard REACLIB library \cite{cyburt_jina_2010}. As such, the theoretical rate calculated with the NON-SMOKER statistical model \cite{rauscher_non-smoker_1999} and found in REACLIB is the standard value used in many stellar models. However, the underlying Hauser-Feshbach formalism used in cases where experimental values are not available, may be inappropriate if the reaction rate is dominated by a small number of resonances in the $^{38}$Ca compound nucleus \cite{rauscher_astrophysical_2000-1}. 

A previous indirect measurement at iThemba Laboratories studied the $^{40}$Ca$(p,t)^{38}$Ca reaction  using the K$=600$ spectrograph \cite{long_indirect_2017}. This measurement confirmed 14 previously observed levels and detected more than 30 new levels between the ground state and E$_x=13$ MeV, with $\sim20$ of these within the Gamow widow for $T_9=1.5$ GK. This result agrees with measurements of the mirror nucleus, $^{38}$Ar, which contains approximately the same number of $s$-wave resonances within this region \cite{Blackmon2014}. These results were used to calculate a new $^{34}$Ar($\alpha,p$)$^{37}$K reaction rate, which is lower by up to two orders of magnitude in the Gamow window when compared with TALYS-1.8 \cite{TALYS} and NON-SMOKER$^{web}$/REACLIB calculations.  However, this discrepancy was not seen in the recent study by Browne \textit{et al.} \cite{Browne_2023}, which used a Si detector array with the Jet Experiments in Nuclear Structure and Astrophysics (JENSA) gas jet target \cite{Chipps2014} for a direct measurement using an $^{34}$Ar beam from ReA3.  Two measurements were taken at $E_{cm}=5.6$ and 5.9 MeV and the resulting cross sections are consistent with the Hauser-Feschbach-calculated value of this rate.  However, that study was not able to disentangle contributions from ($\alpha,p$) reactions on the beam contaminants and the data obtained was above the astrophysically relevent energy regime of $E_{cm}=2.0-3.9$ MeV.

Here we report on a measurement of proton scattering on $^{37}$K (Sec. \ref{exp}) to study properties of states in the compound nucleus, $^{38}$Ca, important for estimating the $^{34}$Ar$(\alpha,p)^{37}$K reaction rate. The excitation spectrum resulting from this experiment and previously known resonance energies were analyzed in an $R$-matrix calculation to constrain spin-parity values and partial widths of these resonances and to identify additional resonances previously unobserved (Sec. \ref{Analysis}). These quantities, together with theoretical $\alpha$ and proton widths, were used to calculate the reaction rate via the narrow-resonance formalism (Sec. \ref{RR}), which is compared to other existing rate calculations. It was then input into models of an XRB in the stellar evolution software MESA (Modules for Experiments in Stellar Astrophysics). The results of these models are then compared with a baseline model and previous models in Sec. \ref{MESA}.
 
\section{Experimental Design}\label{exp}
The experiment was performed at the ReA3 facility at the National Superconducting Cyclotron Laboratory (NSCL) at Michigan State University  using a radioactive ion beam of $^{37}$K ($t_{1/2}=1.23$ s).  A primary beam of $^{40}$Ca from the K500 \& K1200 coupled cyclotrons bombarded a $^{9}$Be production target. The A1900 magnetic fragment separator was used to select $^{37}$K from the resulting isotopes. A gas catcher then collected these ions and fed them into the Electron Beam Ion Trap charge breeder to produce $^{37}$K$^{2+}$ ions. These were injected into the ReA3 linear accelerator to produce a final beam of 4.448 MeV/u with $\sim87\%$ purity and an intensity of $\sim10^{4}$ pps delivered to the experimental area.

\begin{figure}
    \centering
    \includegraphics[scale=.30]{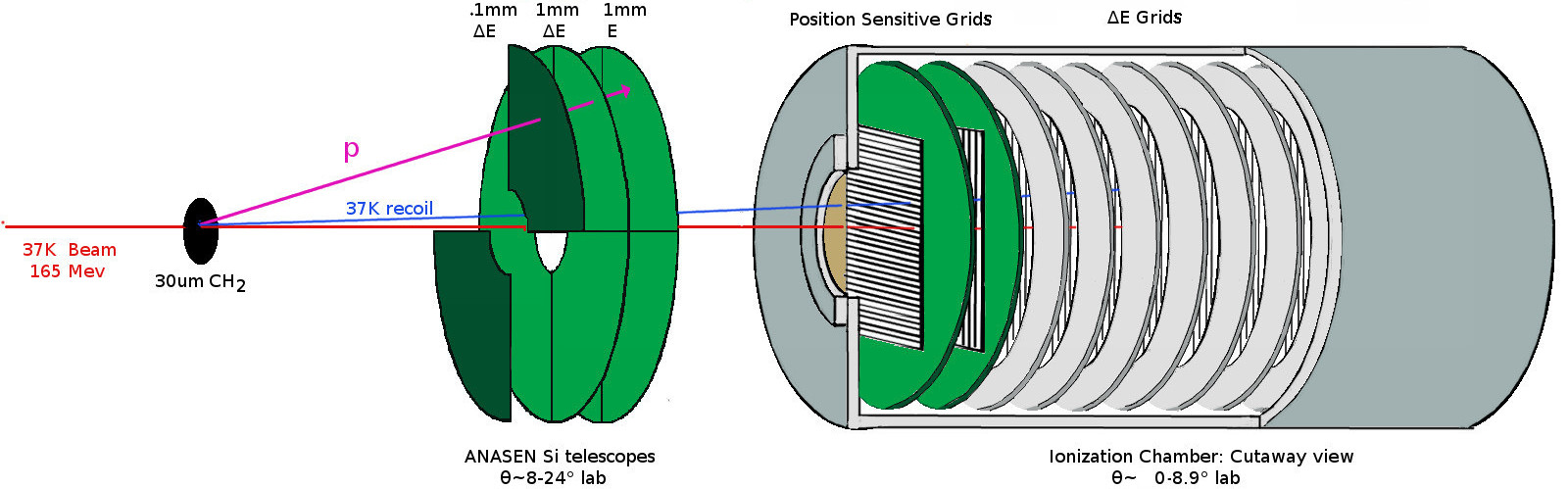}
    \caption{Schematic design of the experimental setup.  The $^{37}$K beam impinges on the polypropylene target, with the lighter product, the protons, emitted at larger angles and entering the silicon telescope.  The heavier products and the unreacted beam are forward focused into the ionization chamber (IC). The green wire grids in the IC are printed circuit boards that provide position sensitivity, while signals from the gray grids are combined in sequence in adjustable groups to measure energy loss and total energy of the particles.}
    \label{fig:detector_system}
\end{figure}

The beam impinged on a 2.7-mg/cm$^{2}$ CH$_2$ target to produce the desired proton-scattering events. Protons emitted at
$\theta_{lab}=15.4-28.2^{\circ}$  were detected by Si detector telescopes arranged in quadrants as shown in Fig. \ref{fig:detector_system}. Two opposite quadrants consisted of a stack of two 1-mm thick Si detectors (both Micron Semiconductor model QQQ3). The other two quadrants comprised a stack of three Si detectors (Micron Semiconductor models QQQ5) with a 0.1-mm thick $\Delta E$ detector backed by two additional Si detectors of 1-mm thickness each.  However, the thresholds in the ASIC electronics for detection of the low energies deposited by the protons in the 0.1-mm-thick detectors were found to be unreliable. Thus, in the subsequent analysis, data is only presented from the QQQ3 detectors. A self-consistent energy calibration was performed by combining the dominant $^{241}$Am characteristic $\alpha$ decays ($E_{\alpha}= 5.486$ MeV) and signals from an electronic pulser.  The heavy ions (beam, contaminants, and heavy recoil products) were forward focused and passed through the center of the silicon telescopes to enter the Ionization Chamber (IC) \cite{Lai_2018}, which provided both position and $Z$ identification by relative energy loss.  The different elemental components of the beam were clearly resolved in the ionization chamber when a target was not in place, allowing the beam composition and intensity to be periodically monitored and optimized. Contaminants in the beam ($\sim$13$\%$ of the beam) were nearly equal parts $^{37}$Cl and $^{37}$Ar. With the CH$_2$ target in place, the $Z$ resolution of the IC was not sufficient to discriminate scattering from these elemental components on an event-by-event basis; however, the scattering cross section for these contaminants is not expected to vary significantly compared to the $^{37}$K$+p$ elastic scattering cross section.

Protons arising from scattering were selected by a gate on the timing coincidence between the IC and the Si telescopes. Background protons produced through  $^{37}$K+$^{12}$C fusion evaporation were highly suppressed as the majority of heavy recoils from this process are stopped in the IC window or dead space (i.e. the small IC volume upstream of the wire grids) due to their high $Z$ and low energy. Additionally, a 3.2-mg/cm$^{2}$ carbon target was inserted to measure the contributions from $^{37}$K+$^{12}$C fusion evaporation, which produced only a small featureless background when the relevant Si-IC timing gate was applied. This background extended to energies higher than the scattered protons, allowing the level of fusion evaporation contamination to also be quantified in the scattering data where it was determined to comprise less than 5\% of the observed events.

\section{Analysis}\label{Analysis}

\begin{figure}
    \centering
    \includegraphics[scale=.4]{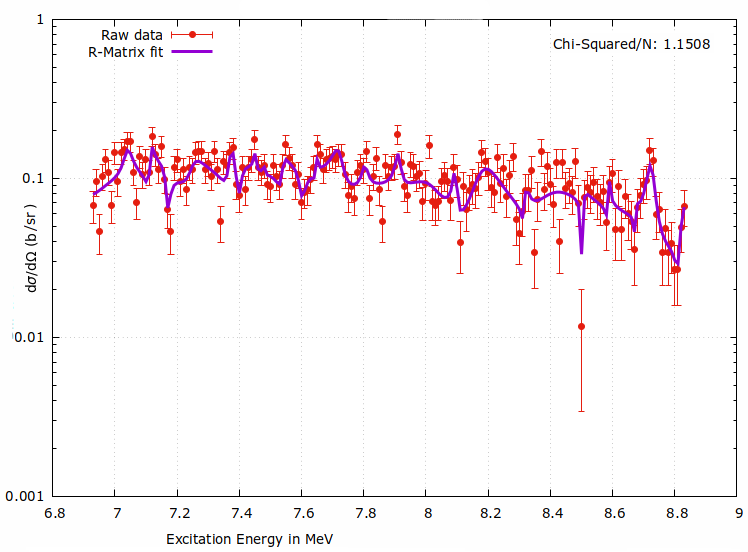}
    \caption{Average differential cross section as a function of excitation energy measured for proton scattering on $^{37}$K in this work.  Also pictured is a representative R-matrix fit calculated at the average angle ($\theta_{cm}=36^{\circ}$) from the process described in Sec. \ref{Analysis} with the corresponding goodness-of-fit statistic, $\chi^2/N=1.1508$.}

    \label{fig:xsec}
\end{figure}

The segmented QQQ3 silicon detectors (16 rings and 16 sectors) measured the reaction angle for each event. Kinematic reconstruction converted the measured proton energy and angle of the elastic scattering events from the lab frame into center-of-mass, combining data over the entire detector into a single excitation function corresponding to the average cross section over the angular range of $\theta_
{cm}=123.4^{\circ}-149.9^{\circ}$.  An absolute energy calibration was applied to the excitation function using proton-scattering data taken with an $^{40}$Ar stable beam and the same CH$_2$ target. This proton-scattering spectrum is well characterized by a previous study \cite{barnard_elastic_1961} with two distinctive resonant features that were aligned with the same features in the $^{40}$Ar stable beam spectrum to provide a two-point calibration. The final, calibrated spectrum from proton scattering on $^{37}$K, converted to excitation energy in $^{38}$Ca in the range of $E_x=6.92-8.82$~MeV (E$_{cm}=0.81-2.71$ MeV), is shown in
Fig. \ref{fig:xsec}.

The excitation function was analyzed using the $R$-matrix formalism \cite{descouvemont_R-matrix_2010} performed with AZURE2 \cite{azuma_azure:_2010}.
A channel radius of $a=6.065$ fm was used, estimated from $R=R_0 (A_1^{1/3}+A_2^{1/3})$ with a suitably large $R_0\sim1.4$ fm to be outside of the range of the nuclear force for the $A=38$ compound nucleus. The excitation function was fit with the calculated differential cross section at a single (average) angle of $\theta_{cm}=136^{\circ}$ to reduce computational complexity, but this is a good approximation  given the limited angular range and modest angular dependence. Previously observed states within the energy range shown in Fig. \ref{fig:xsec} were included with their energy values held fixed to those reported in \cite{long_indirect_2017}.  The success of an AZURE2 fit was judged by the $\chi^2/N$, trying to minimize this value, while reflecting the impact of the resonance energy, partial width, and spin-parity values. 

A good fit to the excitation function could not be obtained using only the levels identified in Long \textit{et al}.  The fit using only those levels resulted in a $\chi^2/N$ of 2.263 and regions of the fit that were not representative of the corresponding features of the experimentally determined cross section.

Only natural-parity states contribute to the $^{34}$Ar$(\alpha,p)^{37}$K reaction due to the $0^+$ total channel spin for the $^{34}$Ar$+\alpha$ system. While population of natural-parity states is favored in the $^{40}$Ca$(p,t)^{38}$Ca reaction \cite{long_indirect_2017}, unnatural parity states can be strongly populated in $^{37}$K$+p$ scattering, so the need for additional states is expected. We added additional hypothetical levels to the $R$-matrix model with energies initially set to the most prominent features in the data. This resulted in a total of 24 states being included in the analysis (13 of which are newly identified here), which is consistent with the level density found in the mirror nucleus $^{38}$Cl. 

With few experimental constraints on resonance properties, the resulting large size of the parameter space and likelihood for many local minima required implementing some assumptions in the analysis.  We included only proton partial widths corresponding to protons populating the ground state in $^{37}$K. The low-energy half of the excitation function ($E_x<7.9$~MeV) may have some contamination from proton inelastic scattering originating from the upstream half of the target that populates the first-excited state in $^{37}$K at 1.37 MeV. However, the observed yield of protons with energies too low to correspond to elastic scattering indicates that this contribution is likely negligible.

Only the lowest orbital angular momentum value was included for any state's particular $J^{\pi}$, but both possible channel-spin combinations were included \cite{DeBoer}. We initially fixed all the resonance energies. Initial values for proton partial widths were chosen that were comparable to the width of features in the excitation function or the experimental resolution from Ref. \cite{long_indirect_2017}, which sets an upper limit on the proton widths for those observed states. The excitation function was fit using MINUIT2 varying partial widths constrained to be within a limited range, varying by no more than a factor of 1.7. Once a local minimum was obtained as demonstrated by the $\chi^2$ within this restricted parameter space, the $J^{\pi}$ value of a randomly selected state was changed, and the gradient fit repeated. The process of randomly changing a $J^{\pi}$ value and repeating a gradient fit was iterated, starting each cycle with the best previous fit, until no improvement was gained.

\begin{center}
\begin{table}
\caption{\label{tab:E_RES} Resonances with suggested spin-parity assignments and partial widths. Resonances not previously identified in \cite{long_indirect_2017} are indicated with *. The systematic uncertainty is indicated with a $^{\diamond}$, while $^{\dagger}$ indicates statistical uncertainties estimated by calculating the fraction of the reduced width amplitude (RWA) uncertainties to the actual RWA value and applying the same ratio to the partial width.
}
\begin{tabular}{|l|c|c|l|l|c|c|}
\hline
\multicolumn{1}{|c|}{\textbf{$E_{x}(MeV)$}} & 
\multicolumn{1}{c|}{\textbf{$J^\pi$}} & 
\multicolumn{1}{c|}{{$\Gamma_p$(keV) ($\pm u^{\diamond}$) ($u^{\dagger}_{+}$)($u^{\dagger}_{-}$)
}}   &
\multicolumn{1}{c}{}&
\multicolumn{1}{|c|}{\textbf{$E_{x}(MeV)$}} & 
\multicolumn{1}{c|}{\textbf{$J^\pi$}} & 
\multicolumn{1}{c|}{{$\Gamma_p$(keV) ($\pm u^{\diamond}$) ($u^{\dagger}_{+}$)($u^{\dagger}_{-}$)
}}   \\




\hline\hline
7.041(8)&$2^+$&33(5)(8)(14)&&7.884(23)$^*$&$1^+/3^+$&21(10)(4)(5)\\
\hline
7.108(11)$^*$&$1^-/1^+$&26(12)(17)(8)&&8.026(5)&$0^+/1^-/2^+$&6(4)(3)(1)\\
\hline
7.176(4)&$3^-$&25(3)(24)(23)&&8.097(3)$^*$&$1^+/2^-/3^-$&12(3)(6)(7)\\
\hline
7.243(8)$^*$&$1^+$&76(33)(24)(19)&&8.132(15)$^*$&$1^+/3^-$&73(2)(70)(36)\\
\hline
7.370(5)&$2^+$&31(7)(8)(7)&&8.189(6)&$2^+$&57(47)(145)(31)\\
\hline
7.411(11)$^*$&$1^+/3^-$&36(7)(43)(72)&&8.322(5)&$2^+$&9(12)(14)(5)\\
\hline
7.450(1)$^*$&$1^+$&5(6)(79)(89)&&8.363(9)$^*$&$1^+$&22(3)(11)(13)\\
\hline
7.539(8)$^*$&$1^+$&36(22)(55)(43)&&8.491(11)$^*$&$1^+$&5(2)(2)(4)\\
\hline
7.611(8)$^*$&$1^+/3^-$&38(20)(26)(72)&&8.507(9)&$1^-$&20(7)(9)(15)\\
\hline
7.647(9)$^*$&$1^+$&45(19)(61)(53)&&8.586(3)&$2^+$&87(78)(14)(13)\\
\hline
7.726(5)$^*$&$3^+$&34(11)(14)(25)&&8.672(6)&$1^-$&8(6)(7)(8)\\
\hline
7.801(3)&$2^+$&28(6)(9)(10)&&8.717(8)&$2^+$&47(7)(14)(14)\\
\hline

\end{tabular}
\end{table}

\end{center}

This entire fitting process was repeated six times, varying $J^{\pi}$ values in different ways each time. An example of one of the fits is shown in Fig. \ref{fig:xsec}.    Finally, fits were performed using the mean values from the results of these six fits as a starting point but varying the energies of previously-unidentified resonances in addition to the partial widths. The resulting fit converged to values close the initial input mean values (well within uncertainties).

Table \ref{tab:E_RES} gives the final values for $E_{x}$, $\Gamma_p$, and $J^{\pi}$. For seven of the 24 states, multiple $J^{\pi}$ values were found to provide a comparable fit. The partial widths given are an average of the six solutions. Uncertainties in the partial widths are provided that include a 
systematic analysis of the values from the six different fits and a statistical uncertainty calculated using the reduced width amplitude values returned from MINUIT2.

\section{Reaction rate}\label{RR}

The reaction rate is calculated using the narrow resonance approximation from the definition provided in \cite{iliadis_nuclear_2007}, with the total reaction rate (in cm$^3$mol$^{-1}$s$^{-1}$) expressed as a sum over individual resonances described by the Breit-Wigner formalism:

\begin{equation}
N_A <\sigma\nu> =1.54 \times 10^{11}(\mu T_9)^{-3/2}\displaystyle\sum_{i}(\omega\gamma)_i \text{exp} \left(-\frac{11.605 E_i}{T_9} \right)
\end{equation}
where $\mu$ is the reduced mass in amu, $T_9$ is the temperature in GK, $E_i$ is the resonance energy in MeV, and $(\omega\gamma)_i$ is defined as the resonance strength (in MeV):
\begin{equation}\label{eq:og}
(\omega\gamma)_i=\frac{(2J_i+1)}{(2j_1+1)(2j_2
+1)}\frac{\Gamma_{\alpha} \Gamma_{p}}{\Gamma_{{tot}}} \approx (2J_i+1)\Gamma_{\alpha}.
\end{equation}
Here, $\Gamma_{\alpha}$ and $\Gamma_{p}$ are the are partial widths of the entrance ($\alpha$) and exit (\textit{p}) channels, respectively; $\Gamma_{{tot}}$ is the total width of the resonance
($\Gamma_{{tot}}=\Gamma_{\alpha}+\Gamma_{p}+\Gamma_{\gamma}$ in this case),
and $J_i$, $j_1$, and $j_2$ are the spins of the resonance and the reactants ($\alpha$ and $^{34}$Ar in this case with $j_1=j_2=0$), respectively.  The $\alpha$ partial width is related to the single-particle width, $\Gamma_{\alpha}^{sp}$, thusly: $\Gamma_{\alpha}$=C$^2S_{\alpha} \Gamma^{sp}_{\alpha}$, where C$^2S_{\alpha}$ is the spectroscopic factor. 
While an $R$-matrix formalism would be more appropriate given the level density, the narrow-resonance approach of \cite{long_indirect_2017} is followed using the same constant $C^2S_{\alpha}=0.01$ to allow direct comparison to the previous work.
Proton partial widths from the $R$-matrix analysis discussed above and estimated gamma partial widths were used to verify that $\Gamma_{p}>>\Gamma_{\gamma}+\Gamma_{\alpha}$, implying resonance strengths can be accurately estimated using $(\omega\gamma)_i\approx (2J_i+1)\Gamma_{\alpha}$.

\begin{figure}
    \centering
    \includegraphics[scale=.65]{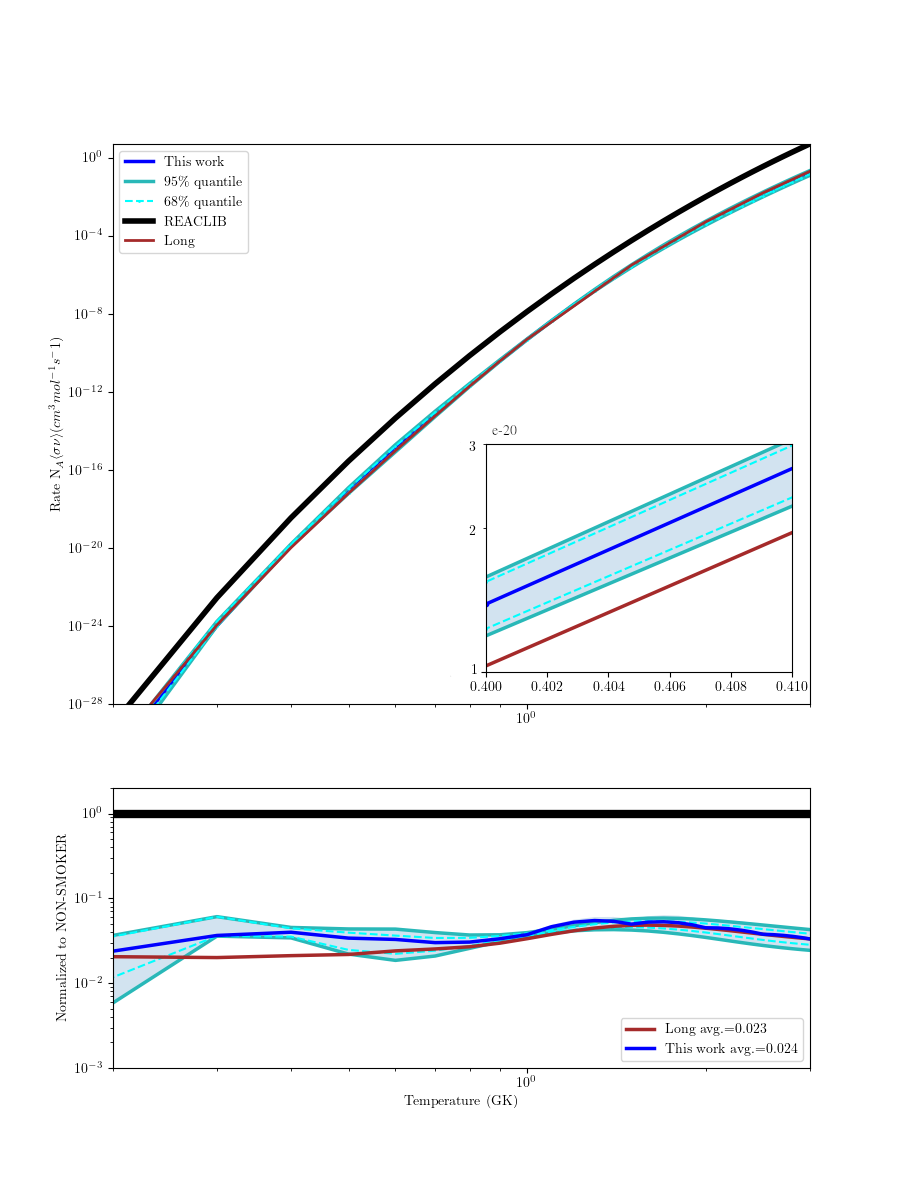}
    \caption{(TOP) The mean astrophysical reaction rate calculated from Monte Carlo sampling is depicted here along with the 68\% and 95\% quantiles. The REACLIB rate is also shown, as well as the rate calculated in Long \textit{et al.} \cite{long_indirect_2017}. (BOTTOM) The ratio of the rates depicted above normalized to REACLIB (i.e. NON-SMOKER).  The gray band represents the range of values resulting from the Monte Carlo sampling.}
    \label{fig:rates}
\end{figure}

The total reaction rate includes the new resonance information from this work as well as additional levels reported in Ref. \cite{long_indirect_2017,Long2016}, as the latter includes levels outside the experimental region explored here that still contribute to the overall reaction rate.  The reaction rate was calculated $10^7$ times using a Monte-Carlo sampling of natural-parity spin assignments for each resonance taken from Long \textit{et al.} (see Fig. 3 in \cite{long_indirect_2017}) outside of our experimentally measured energy region.  For states between E$_x=7-8.7$ MeV, the lowest natural parity assignment from Table \ref{tab:E_RES} was used or the state was not included if the only available spin assignment(s) was of unnatural parity.  The median rate, which is the sum of these contributions, with the 1$^{st}$ and 3$^{rd}$ quartile of rates are presented in Fig. \ref{fig:rates} along with the rates from the previous work by \cite{long_indirect_2017,Long2016}. We also include the rate from REACLIB \cite{cyburt_jina_2010} based on NON-SMOKER \cite{rauscher_non-smoker_1999}, which is nearly the same as the statistical model rate from TALYS-1.8 \cite{TALYS} using the default model parameters.

Our rate is higher than the rate of Long \textit{et al.} \cite{long_indirect_2017} for $T<0.6$ GK due to the inclusion of additional levels but generally consistent at higher temperatures. However, compared to the statistical model rates, our rate is lower by a factor of $\sim20-40$. Given that the density of states in the region is close to that of the mirror nucleus, it is unlikely that a large number of states are missing. The simple assumption of $C^2S_{\alpha}=0.01$ may lead to an underestimate of the rate, particularly as $\alpha$-cluster states could result in significantly higher $C^2S_{\alpha}$, though likely less than the observed discrepancy with the statistical model predictions \cite{long_indirect_2017}. 
Direct measurement of the $^{34}$Ar($\alpha,p$)$^{37}$K total cross section at higher energies found good agreement with statistical model predictions in the regime where the reaction proceeds to many excited final states. However, population of the ground state in $^{37}$K was found to be much weaker than model predictions \cite{Browne_2023}. Population of the ground state is expected to dominate at the energies covered in this measurement (and much of the Gamow window), and indeed is the only state that can be populated in the lower part of the Gamow window.  Thus, the reliability of the statistical model in this lower energy regime is less clear, and uncertainties in the reaction rate are still substantial.

\section{Impact on Stellar Models}\label{MESA}
Given the uncertainty in the rate, we re-examined the impact of the $^{34}$Ar($\alpha,p$)$^{37}$K reaction rate using a Type-I X-ray Burst model adapted from Paxton \textit{et al.} \cite{Paxton2015}, Modules for Experiments with Stellar Astrophysics (MESA), version 10108. MESA is a multi-zone, 1D radial stellar evolution code, with adaptive mesh in time and space that includes relevant physics such as opacities, hydrodynamics, mixing length theory, atmospheric boundary conditions, and equation of state. The nuclear network differential equations are coupled to those that model the physical aspects so that the energy generation provides feedback into the thermodynamics and vice versa, and includes mixing between zones. The model is designed to simulate conditions of a hydrogen-dominant burst on GS 1826-24, colloquially known as the ``clocked burster" due to its extreme regularity \cite{Galloway_2004}, with an initial neutron-star mass and radius of 1.4 M$_\odot$ and 11.2 km, respectively. This model assumes solar abundances for accretion ($^1$H $=0.7048$, $^{4}$He $=0.2752$, and $Z$ from \cite{Grevesse1998}) at a rate of $\sim3\times10^{-9}$ M$_\odot/$yr.  The nuclear network is similar to the classic 304-species XRB network identified by \cite{fisker_explosive_2008}, favoring proton-rich isotopes up to $^{107}$Te; however, the base layer down to the degenerate core is modeled as the heavy element $^{138}$Ba, which is technically included in the nuclear network but is far above the next-greatest $Z$ isotope and so unreachable by any standard capture reaction, making the core inert.  Additional astrophysical details of the model can be found in Ref. \cite{Paxton2015}.

Here we report the results of 11 variations of the base model to test a sample of the range of known burst conditions with different $^{34}$Ar($\alpha,p$)$^{37}$K reaction rates. For five models there was no alteration save the $^{34}$Ar($\alpha,p$)$^{37}$K reaction rate. This set included the standard REACLIB rate, REACLIB $\times100$ and $\times0.01$, the recommended rate from this work, and REACLIB $\times0.022$ (i.e. approximately the average ratio of this work to REACLIB). These factors were chosen to compare with similar published stellar modeling studies. The standard REACLIB rate is from the 2010 compilation of theoretical Hauser-Feschbach rate calculations \cite{cyburt_jina_2010}. The MESA models were run for 3000 timesteps and all analysis values were calculated over these, except for abundances which were calculated after quiescence at the 10$^{th}$ burst.

Three additional models increased the He accretion percentage to $47\%$ to mimic a He-rich companion and provide conditions where $\alpha$-induced reactions might be especially impactful.  Finally, three more models tested a slowed accretion ($7.93\times10^{-10}$M$_{\odot}$/yr) compared to the ``clocked burster" model.  These six models all used the $^{34}$Ar($\alpha,p$)$^{37}$K REACLIB rate as the baseline with the same variations by factors of $\times100$ and $\times0.01$.

\begin{figure}
    \centering
    \includegraphics[scale=.5
]{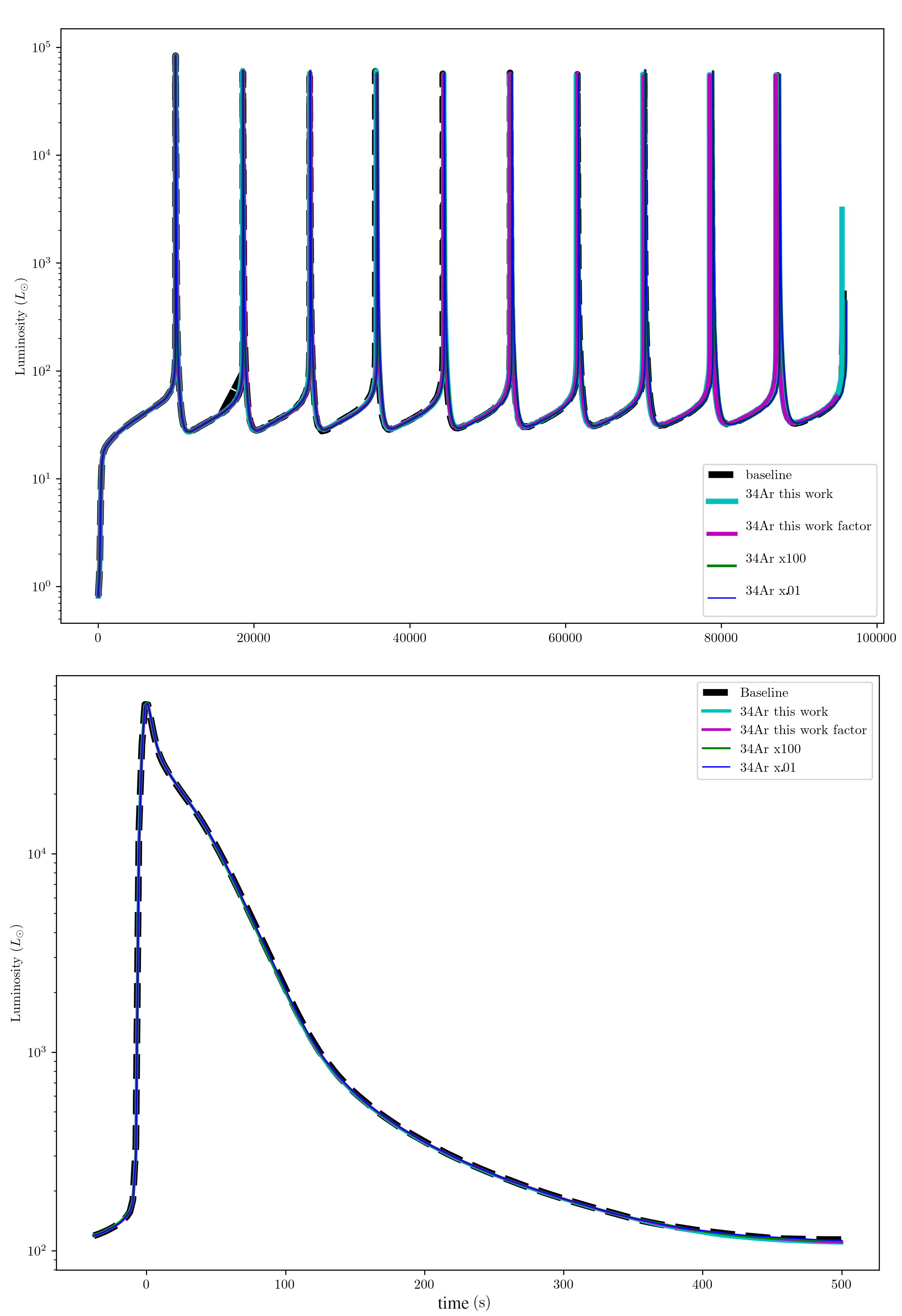}
    \caption{(TOP) The full series of bursts for the ``clocked burster" scenario and accompanying rate variations. (BOTTOM) The luminosity results of each model are averaged and ``folded", such that their peaks are aligned. The resulting plot demonstrates that the variation in burst frequency for the folded light curve caused by changes in reaction rate is not significant enough to be detected on visual inspection. However, variations in the magnitude of the luminosity may explain integrated luminosity variations. Numerical values describing these results are available in Table \ref{tab:model_res}.  Note "factor'' refers to the 0.022 factor described in the text.}
    \label{fig:folded}
\end{figure}

\begin{figure}
    \centering
    \includegraphics[scale=.45]{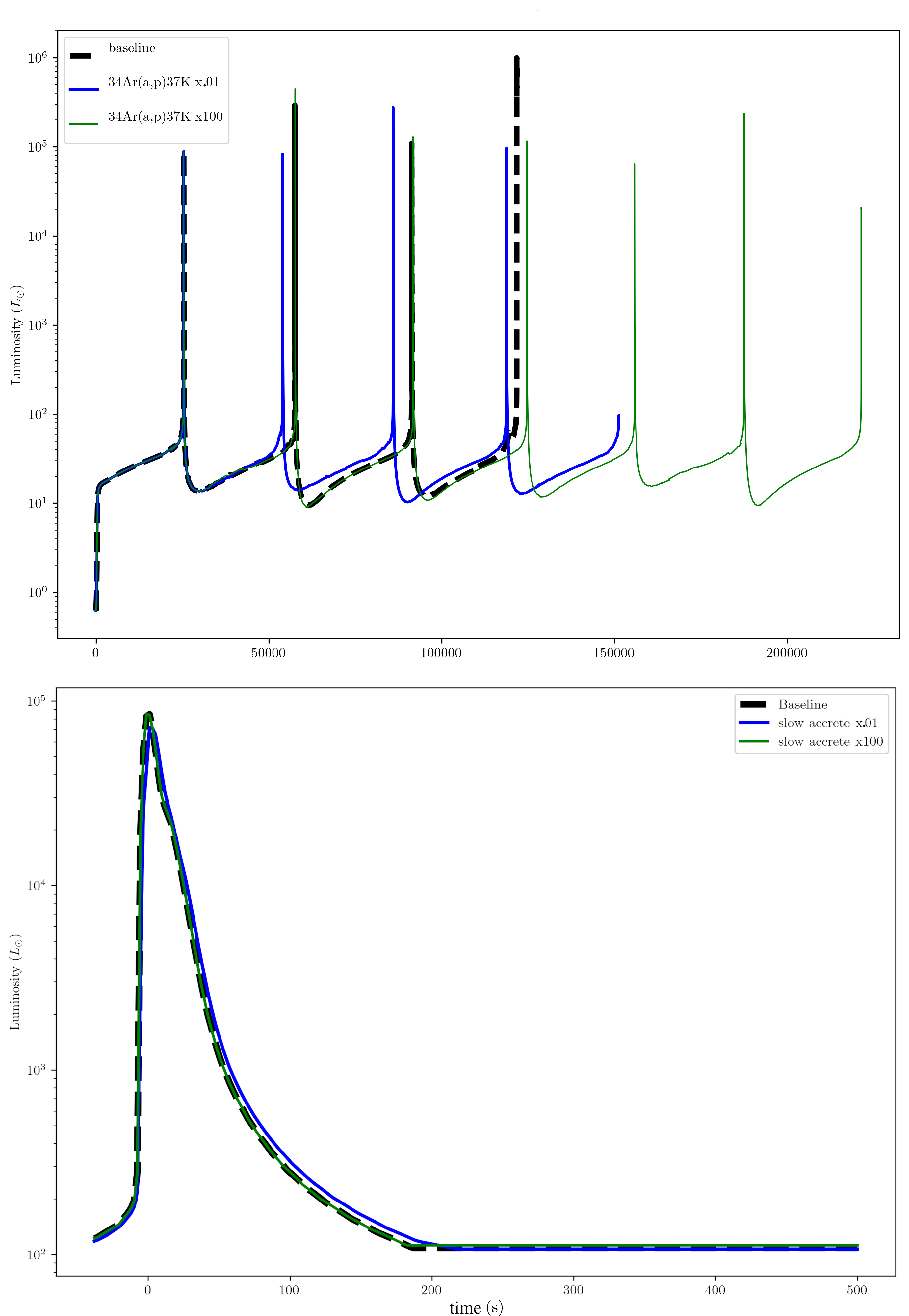}
    \caption{(TOP) The full series of bursts for the slow accretion scenario and accompanying rate variations. (BOTTOM) The same set of models ``folded" such that they are averaged for a given model and their peaks aligned.}
    \label{fig:Model_SA}
\end{figure}

\begin{figure}
    \centering
    \includegraphics[scale=.45]{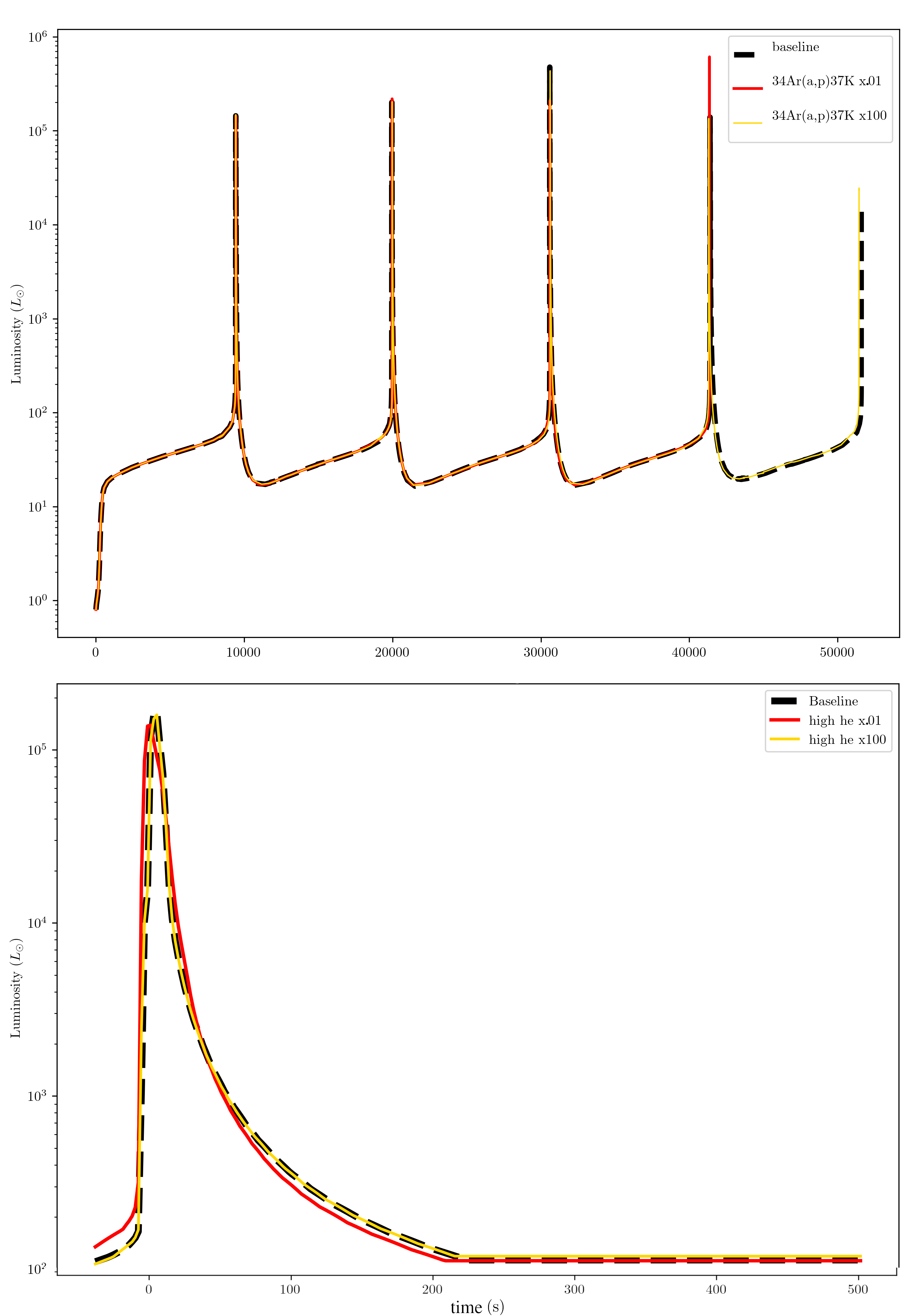}
    \caption{(TOP) The full series of bursts for the high helium accretion scenario and accompanying rate variations. (BOTTOM) The same set of models ``folded" such that they are averaged for a given model and their peaks aligned. Numerical values describing these results are available in Table \ref{tab:model_res}.}
    \label{fig:Model_HH}
    \end{figure}

The models investigated here were compared via the following metrics: burst period (absolute and differential), integrated luminosity difference (definition given below), and maximum luminosity (absolute and differential). A summary of the results of the stellar modelling is given in Table \ref{tab:model_res}. The number of bursts is also listed, and gives a general sense of the numerical stability of the model, with more bursts corresponding to higher stability.  The full set of stable burst models as well as the folded, averaged burst light curves, which are used to calculate integrated luminosity metric, are reflected in Figs. \ref{fig:folded}, \ref{fig:Model_SA}, and \ref{fig:Model_HH}.

The three different sets of models result in differences in the burst frequency from each other, but as this was not caused by variations in the $^{34}$Ar($\alpha,p$)$^{37}$K rate, it is not relevant to this work. The only significant change in burst frequency due to this rate variation was in the model using slower accretion (see Fig. \ref{fig:Model_SA}) with the $^{34}$Ar($\alpha,p$)$^{37}$K rate set at $\times0.01$ the standard REACLIB rate, which was 2689 s (9$\%$) longer. The same model with the rate varied by $\times100$ showed a slightly smaller variation, while other models varied by $<1\%$.  The absolute maximum luminosity showed more significant variations in the models tested, with the largest variation of 260\% again in the slower accretion model with the reaction rate set at $\times0.01$ the REACLIB rate, followed by the slow accretion model with reaction rate at $\times100$ (120\%), and the high He-accretion models with the rate at $\times0.01$ (22\%) and $\times100$ (9\%).

The integrated burst luminosity is a metric designed to determine the overall effect on the light curve when a reaction rate is varied, defined in \cite{cyburt_dependence_2016} as

\begin{equation}
   M_{LC}^{(i)} = \int_{-10}^{150} {|\langle L_{i}(t)\rangle - \langle L_0(t)\rangle} | dt,
\end{equation}

\noindent where $L_0$ is a baseline model with accepted rates from REACLIB \cite{cyburt_jina_2010}. This quantity was similar for all of the ``clocked burster" variations, which differed by less than 13\% between models. Though these models had similar burst periods and maximum luminosities, variations in burst and quiescent period shape and values could explain these non-zero values (i.e. the differences from the baseline model). The high He-accretion model using the standard reaction rate $\times0.01$ had the largest $M_{LC}^{(i)}$ value, likely due to the larger difference in maximum luminosity compared with the Clocked-Burster series, along with more variation throughout the burst tail. Both slow accretion models had modest values for this metric (Table \ref{tab:model_res}).

\begin{center}  
\begin{table}
\caption{\label{tab:model_res}Summary of the numerical results of the metrics described in Section \ref{MESA}.}
\begin{tabular}{|l|r|r|r|r|r|r|}
\hline
Model Identifier & no. bursts &  burst period (s) &  burst diff (\%) & $M^{(i)}_{LC} (10^{38}$ erg) &  $L_{max} (10^{38}$ erg/s) &  $L_{max}$ diff ($\%$) \\
\hline\hline
Clocked burster & \multicolumn{6}{c|}{}\\
\hline
REACLIB rate &          11 &       8577.07 &         0.00 &             0.00 &            83258.96 &          0.00 \\
this work &          10 &       8566.84 &         0.12 &             6.89 &            83123.07 &         -0.16 \\
REACLIB $\times0.022$ &           9 &       8576.47 &         0.01 &             7.74 &            83303.02 &          0.05 \\
REACLIB $\times100$ &          10 &       8622.79 &        -0.53 &             7.38 &            83657.11 &          0.48 \\
REACLIB $\times0.01$ &           9 &       8601.35 &        -0.28 &             7.82 &            83113.71 &         -0.17 \\
\hline
high He fraction & \multicolumn{6}{c|}{}\\
\hline
REACLIB rate &           4 &      10649.18 &         0.00 &             0.00 &           474962.17 &          0.00 \\
REACLIB $\times0.01$ &           3 &      10579.34 &         0.66 &             9.95 &           611728.20 &         22.36 \\
REACLIB $\times100$ &           4 &      10619.49 &         0.28 &             1.13 &           434371.27 &         -9.34 \\
\hline
slow accretion rate & \multicolumn{6}{c|}{}\\
\hline
REACLIB rate&           3 &      32957.84 &         0.00 &             0.00 &           993079.72 &          0.00 \\
REACLIB $\times0.01$ &           3 &      30268.30 &         8.89 &             5.08 &           276449.56 &       -259.23 \\
REACLIB $\times100$ &           6 &      32394.70 &         1.74 &             3.22 &           447802.64 &       -121.77 \\
\hline

\end{tabular}
\end{table}
\end{center}

\section{Conclusions}
Resonances of $^{37}$K$+p$ in $^{38}$Ca have been studied via proton scattering with a $^{37}$K beam on a CH$_2$ target at the ReA3 facility using a Si detector array and a position-sensitive ionization chamber to determine the excitation function between approximately $E_x=7.0 - 8.7$ MeV in the $^{38}$Ca compound nucleus.  In addition to the identification of 13 new levels, the J$^{\pi}$ values and proton partial widths of these levels, as well as for levels previously identified by Long \textit{et al.} \cite{long_indirect_2017} in this energy range, have been constrained through an R-Matrix analysis of these results. This information was used to estimate the $^{34}$Ar($\alpha,p$)$^{37}$K reaction rate, which was found to be approximately a factor of 20-40 less than the statistical model rates that have been shown to describe the cross section at higher energies.

The sensitivity of XRB models to the $^{34}$Ar($\alpha,p$)$^{37}$K rate was studied using MESA astrophysical simulations based on GS 1826-24.  Outputs of a baseline model run with the current recommended reaction rate in REACLIB were compared to those with variations of that rate by factors  of  100 up and down, as well as the rate determined in this work.  
We find no significant deviations in the luminosity profile, burst recurrence time, or burst duration of this model when the $^{34}$Ar($\alpha,p$)$^{37}$K rate is varied; however, large variations in the maximum luminosity due to this rate were observed for models with slower accretion rates or a larger He fraction.  This does not agree with the results from the single zone models described in Cyburt \textit{et al.} \cite{cyburt_dependence_2016} or Browne \textit{et al.} \cite{Browne_2023}, nor with the results found in the KEPLER multizone model of Cyburt \textit{et al.} when the $^{34}$Ar($\alpha,p$)$^{37}$K rate was varied by the same factor of 100 as in this work.  Such differences between multizone and single zone models have been observed previously \cite{cyburt_dependence_2016} and are not surprising, while the differences between the MESA models presented here and the multizone models of Cyburt \textit{et al.} may be due to the fact that this reaction is near the endpoint of the $\alpha,p$ process and as such small differences in the temperature profile as a function of depth may have large effects on the models.  It is possible these results could serve as caution against relying heavily on post-processed single-zone models, especially when the nuclear network includes no feedback with the thermodynamics, as is often the case (although \cite{cyburt_dependence_2016} did include feedback in their one zone model).  This also suggests that the sensitivity of various models to different reactions can vary based on the model's parameters, architecture, etc.,  especially given the vast improvement in computational capabilities over the intervening years.

\section{Acknowledgments}
This work was partially supported by the U.S. Department of Energy, Office of Science contract no. DE-SC0014231 and Office of Nuclear Physics under contract numbers DE-FG02-96ER40978, DOE DE-FG02-02ER41220, DE-AC05-00OR22725, DE-FG02-93ER40773, and DE-AC02-06CH11357, Louisiana Board of Regents RCS Subprogram Contract LEQSF(2012-15)-RD-A-07, the National Science Foundation under contract number PHY-1401574, and IBS grant number IBS-R031-D1, funded by the Korea government.

\bibliography{alaueraps}

\end{document}